\documentclass[preprint,fleqn,showpacs,showkeys]{revtex4}
\usepackage{amsfonts}
\usepackage{amsmath}
\usepackage{amssymb}
\usepackage{graphicx}
\begin{document}
\setcounter{page}{0}
\title[The stability of 3 transmembrane and 4 transmembrane human vitamin K epoxide reductase models]{The stability of 3 transmembrane and 4 transmembrane human vitamin K epoxide reductase models}
\author{Sangwook \surname{Wu}}
\email{sangwoow@pknu.ac.kr}
\thanks{Fax: +82-51-629-5549}
\affiliation{Department of Physics, Pukyong National University,
  Busan, 608-737, Korea}


\begin{abstract}
3 transmembrane and 4 transmembrane helices models are suggested for the human vitamin K epoxide reductase (VKOR). In this study, we investigate the stability of the human 3 transmembrane/4 transmembrane VKOR models employing a coarse-grained normal mode analysis and molecular dynamics simulation. Based on the analysis of the mobility of each transmembrane domain, we suggest that the 3 transmembrane human VKOR model is

more stable than the 4 transmembrane human VKOR model.
\end{abstract}



\maketitle

\section{INTRODUCTION}
Human vitamin K epoxide reductase (VKOR) plays a critical role in generating gamma-carboxyglutamate, one of the essential ingredient in blood coagulation cascade \cite{Furie,Sadler}. The amino acid sequence of human VKOR enzyme (163-residues) was identified \cite{stafford,oldenburg}. Many biochemical experiments have been performed on the study of the human VKOR enzyme. However, the X-ray crystal structure of the human VKOR enzyme is not known yet. In the absence of the X-ray crystal structure, two models for the topology of the human VKOR enzyme have been suggested: 3 transmembrane(TM) \cite{Wu2}/4 transmembrane(TM) \cite{oldenburg2,timson} VKOR models. For the human 4TM VKOR model, it is derived from the bacterial VKOR homolog \cite{Beckwith}. The initial TM domains of the human 4TM VKOR are defined as TM1: R12-A31, TM2:S81-L97, TM3:L105-V127, and TM4:C132-R151. On the other hand, for the human 3TM VKOR model, it is determined based on the biochemical experiments \cite{Tie1,Tie2,Tie3} and computer-aided program \texttt{TOPCONS} \cite{sangwook,Topcons}.
A schematic topology for the 3TM/4TM human VKOR models. The topology difference between the two models make the active sites, C132 and C135 (colored as yellow), locate in different positions within the lipid bilayer.
The definitions of the TM domains for the human 3TM VKOR are determined as TM1: (G9-V29), TM2: (V104-L124), and TM3 (V127-W147).
The most significant difference in the topology of the human 3TM/4TM VKOR models is that the N-terminus of the human 3TM VKOR model starts from the lumen while the C-terminus ends at the cytoplasm. On the other hand, for the human 4TM VKOR model, the N-terminus starts from the cytoplasm and the C-terminus ends at the cytoplasm. Figure 1 shows a schematic topology for the human 3TM/4TM VKOR model. The last two TM domains (2nd and 3rd TM domains in the human 3TM VKOR model and 3rd and 4th TM domains in the human 4TM VKOR model) show common feature in their membrane topology.
The difference between the topology of the human 3TM/4TM VKOR models makes the two active sites, C132 and C135, locate in different positions inside the lipid bilayer.\\
Normal Mode Analysis (NMA) is an effective tool for analyzing the slowest motions of protein \cite{Levitt,Marques,Hinsen}.
\begin{figure}
	\begin{centering}
	\includegraphics[width=0.4\textwidth]{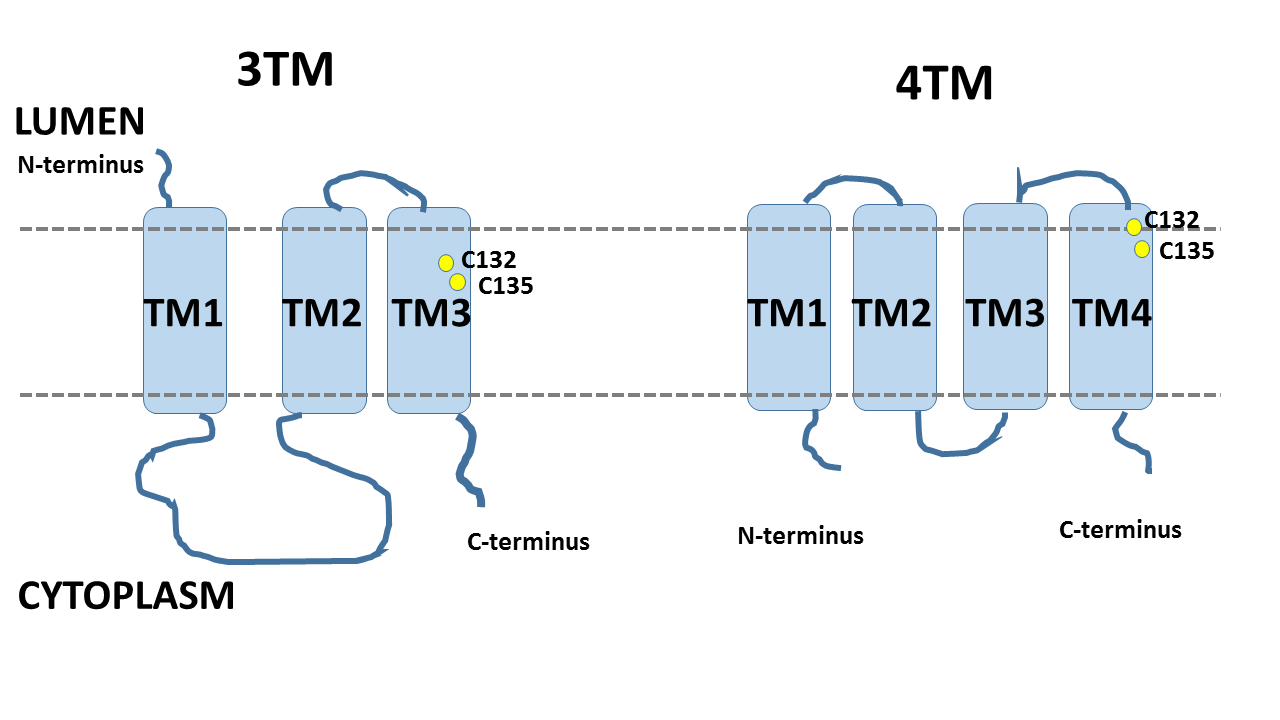}
	\caption{A schematic topology for the 3TM/4TM human VKOR models. The topology difference between the two models make the active sites, C132 and C135 (colored as yellow), locate in different positions within the lipid bilayer.}
	\label{fig:1 A schematic topology}
	\end{centering}
\end{figure}
\begin{figure}
	\begin{centering}
	\includegraphics[width=0.5\textwidth]{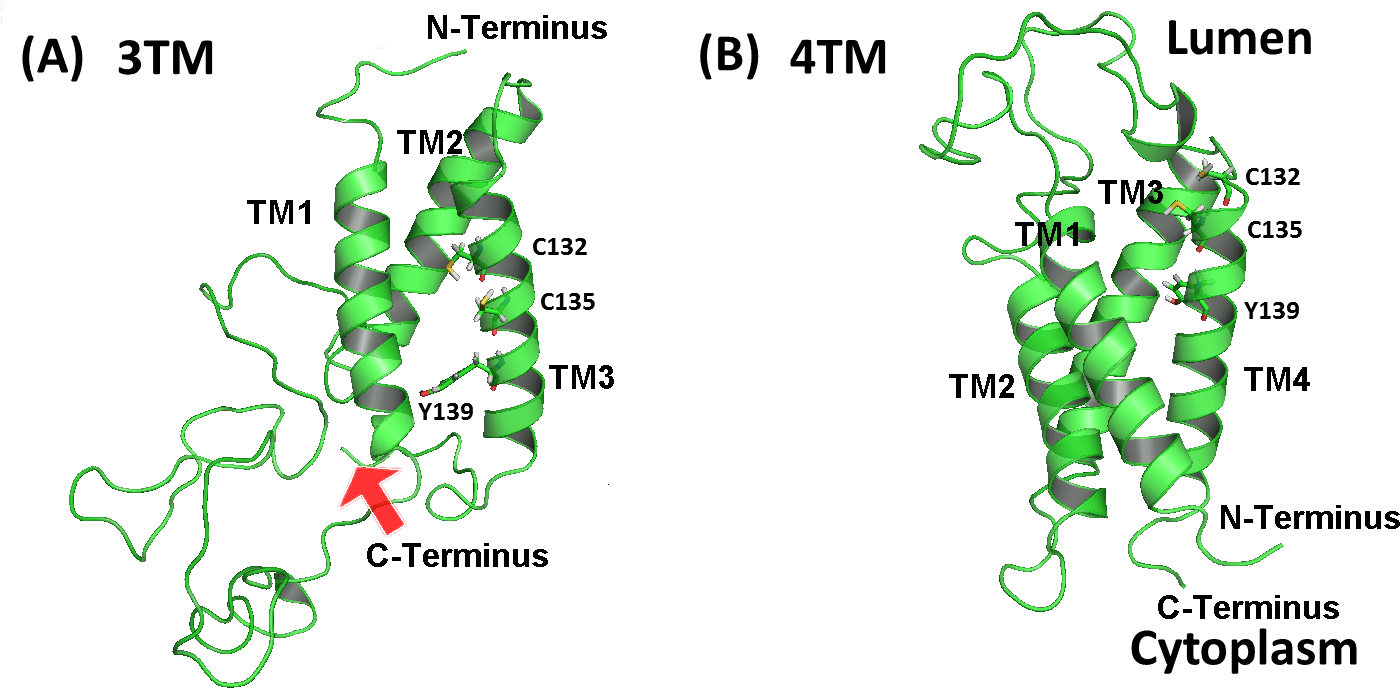}
	\caption{(A) 3-dimensional 3TM human VKOR model (B) 3-dimensional 4TM human VKOR. Two active sites, C132 and C135, are shown. Y139, one of the proposed warfarin binding residues, is shown. For clarity, the lipid and water molecules are not shown. The image is modified from the Ref.5}
	\label{fig:2 real topology}
	\end{centering}
\end{figure}
For the system in the equilibrium, the potential energy V can be expanded as Taylor series \cite{Reuter2},
\begin{equation}
\mathbf{V}=\frac{1}{2}\left( \frac{\partial ^{2}V}{\partial q_{i}\partial
q_{j}}\right) _{0}\eta _{i}\eta _{j}
\end{equation}
where $\eta$ is a displacement from the equilibrium position, $q$=$q_{0}$+$\eta$. 0 means the equilibrium position, $q_{0}$ in the generalized coordinate.
For the oscillating system, the deviation can be expressed as
\begin{equation}
\eta _{i}=\sum_{k}a_{ik}\cos (\omega _{k}t+\delta _{k})
\end{equation}
where $\omega_{k}$ is the angular frequency and $\delta_{k}$ is the phase. And $a_{ik}$ is the component of the amplitude matrix $\textbf{A}$.
Then, we can solve the equation in the matrix form.
\begin{equation}
\text{A}^{\text{T}}\text{VA}=\mathbf{\lambda}
\end{equation}
\textbf{A} is the matrix of the oscillation amplitude.\textbf{V} is the matrix of the second derivatives of the potential energy, Hessian. \textbf{$\lambda$} is the eigenvalue matrix. The eigenvectors of the Hessian correspond to normal modes. For accurate normal mode analysis for protein, it requires a full potential for all the atoms of the protein. However, it requires an expensive computational cost.
To save the computational time, coarse-grained normal mode analysis (dealing with $C_{\alpha}$ atoms) was developed \cite{Hinsen}.
The harmonic pair potential of the coarse-grained model, based on the elastic network model (ENM) \cite{tirion}, has a form
\begin{equation}
V(r)=\frac{1}{2}\sum_{ij}k(\left\vert R_{ij}^{0}\right\vert )(\left\vert
r\right\vert -\left\vert R_{ij}^{0}\right\vert )^{2}
\end{equation}
where $R_{ij}^{0}$ is the pair distance vector. And the spring constant k(r), function of the atomic pair distance, is obtained from the fitting to the Amber94 force field \cite{Hinsen2}.
In this study, we analyze the mobility of TM helices of the two proposed models for the human VKOR using the normal mode analysis and molecular dynamics simulation.
\section{METHODS}
\subsection{Normal Mode Analysis}
The translational movement along the axis, the rotational movement around the axis, the slide movement perpendicular to the axis towards/away from the center of the bundle, and the tilt perpendicular to the axis away from the center of bundle are defined in the web-application \texttt{TMM@server} \cite{Reuter} (insets in Figure 3). The axis of $\alpha$-helix is defined as the principal axis of inertia of $C_{\alpha}$ atoms. The projections are defined as \cite{Reuter}
\begin{equation}
p_{i}=\text{d}\cdot e_{i}
\end{equation}
where d is the displacement vector including the rotation, translation, slide and tilt. $e_{i}$ is the normal mode vector of i. N is the number of atom. Cumulative squared overlap calculates how much normal mode i contribute to the displacement. The range of the cumulative squared overlap is 0 to 1 with the relation of \cite{Reuter}
\begin{equation}
\sum_{i=1}^{3N}p_{i}^{2}=1.
\end{equation}
Mode numbers are shown in x-axis as vibrational frequencies of movements. In the cumulative squared overlap graph, a curve reaches an asymptotic behavior for fewer vibrational modes means greater mobility of TM domain \cite{Reuter}.
\subsection{Molecular Dynamics Simulation}
We performed molecular dynamics simulation of the \emph{in-silico} human 3TM/4TM VKOR model \cite{Wu2}. We employed the NAMD 2.8 package \cite{NAMD} with the CHARMM 27 force field \cite{CHARMM} with the protein parameters incorporating the CMAP terms \cite{Wu2,cmap}. A palmitoyl-oleoyl-phosphatidylethanolamine lipid bilayer (POPE) was prepared with the dimension of 100 \AA $\times$ 100 \AA  with the TIP3P water model \cite{TIP3P}. For physiological condition, we set the system to be 150 mM in NaCl environment. We performed an energy minimization over 20,000 steps by the conjugate gradient method. Subsequently, the system was heated to 310 K over 60 ps. The particle mesh Ewald (PME) method \cite{PME} was used. Molecular dynamics simulation in the NPT ensemble (310 K, 1 atm) was performed for over 2 ns. Constant pressure (1atm) was maintained by using the Langevin piston Nose-Hoover method \cite{Hoover}. The MD simulation scheme was changed to the NPnTA ensemble for 60 ns simulation.
\section{RESULT AND DISCUSSION}
\begin{figure}
	\begin{centering}
	\includegraphics[width=0.6\textwidth]{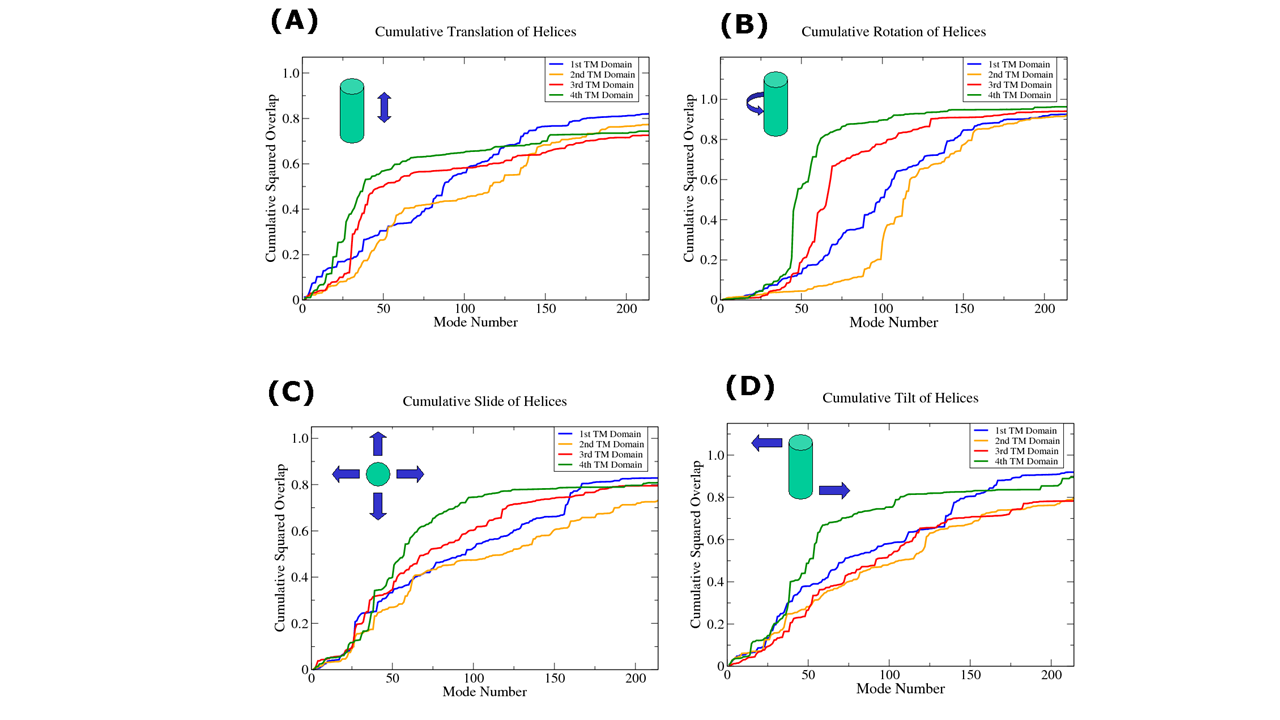}
	\caption{(A) translational movement along the axis (B) rotational movement around axis (C) slide movement perpendicular to the axis towards/away from the center of the bundle (D) tilt perpendicular to the axis away from the center of bundle for the 4TM human VKOR model.}
	\label{fig:3 normal mode analysis 1}
	\end{centering}
\end{figure}
\begin{figure}
\begin{centering}
	\includegraphics[width=0.6\textwidth]{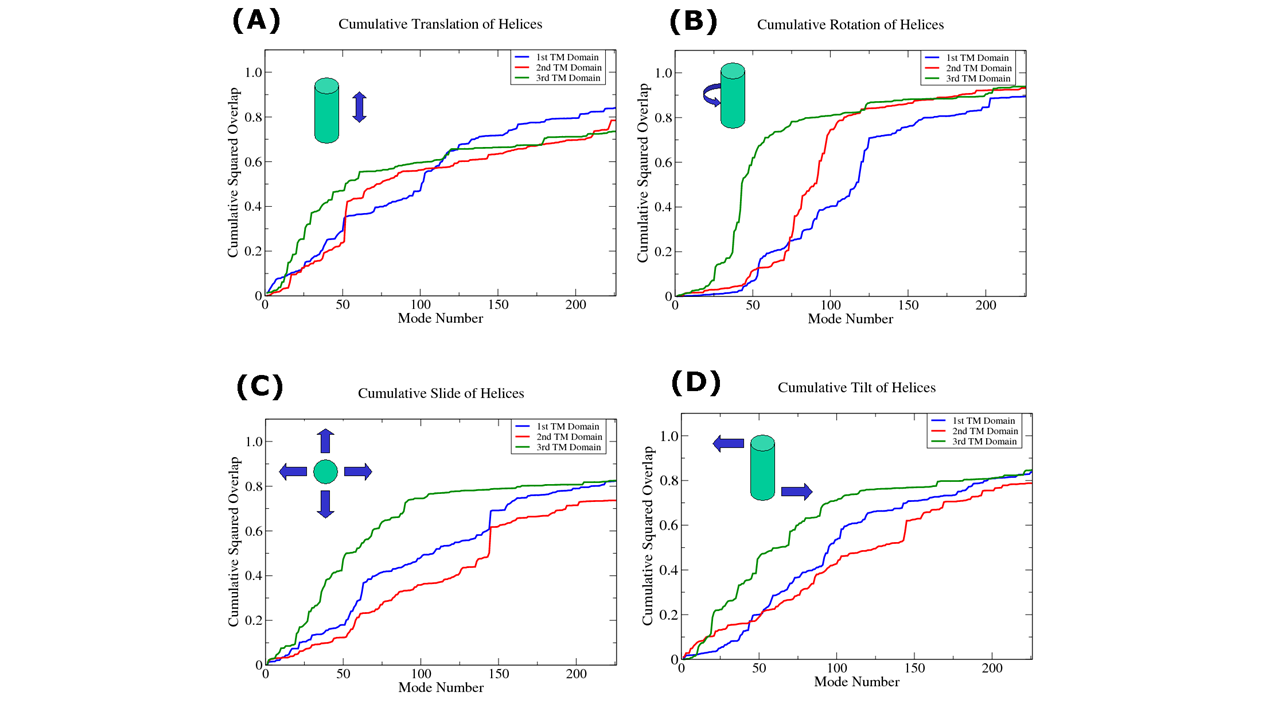}
	\caption{(A) translational movement along the axis (B) rotational movement around axis (C) slide movement perpendicular to the axis towards/away from the center of the bundle(D) tilt perpendicular to the axis away from the center of bundle for the 3TM human VKOR model.}
	\label{fig:4 normal mode analysis 2}
	\end{centering}
\end{figure}
\begin{figure}
	\begin{centering}
	\includegraphics[width=0.6\textwidth]{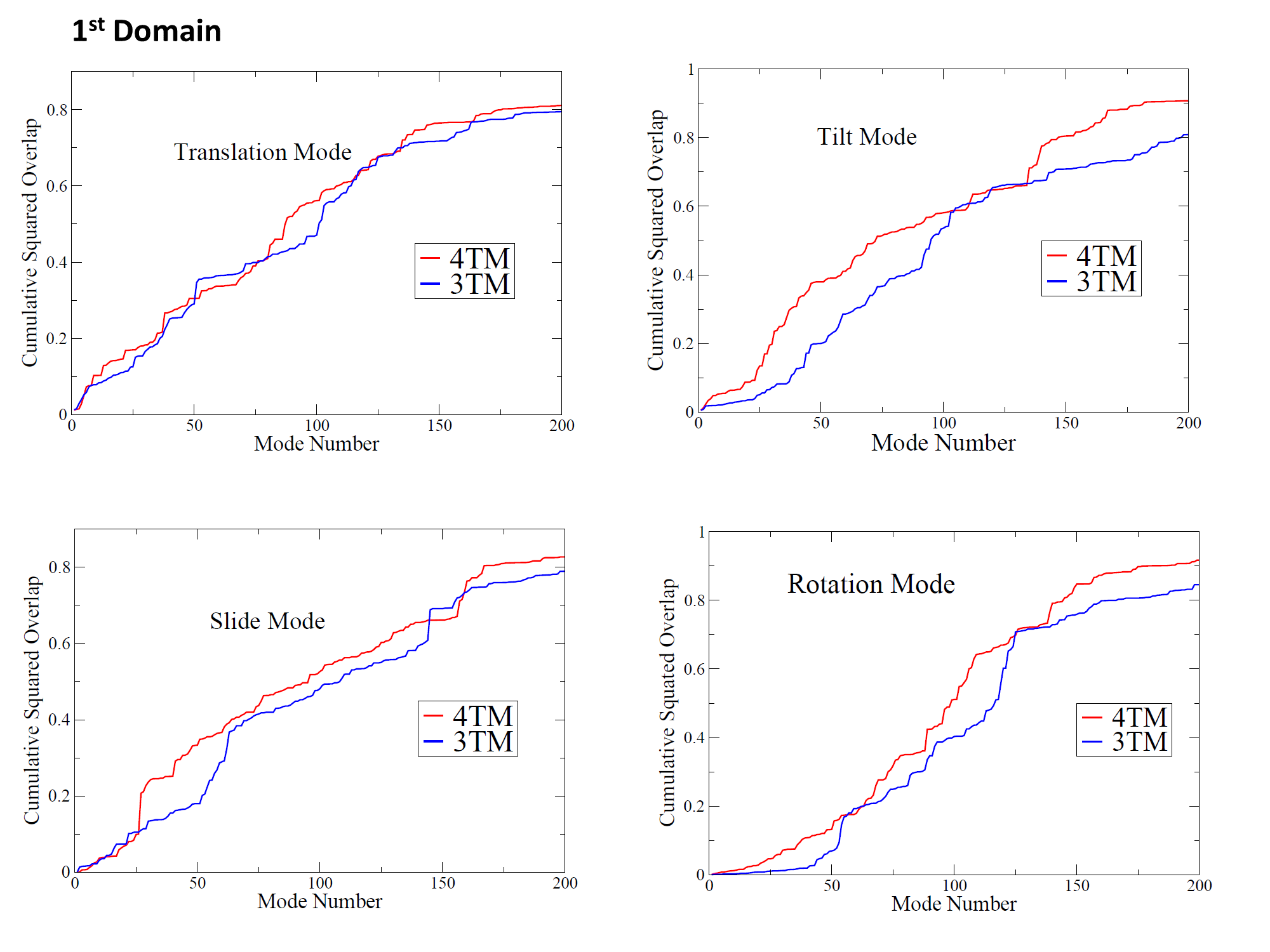}
	\caption{The cumulative squared overlap values for the 1st domain of the 3TM/4TM human VKOR models in translational movement along the axis, rotational movement around axis, slide movement perpendicular to the axis towards/away from the center of the bundle, and tilt perpendicular to the axis away from the center of bundle. }
	\label{fig:5 normal mode analysis 3}
	\end{centering}
\end{figure}
\begin{figure}
	\begin{centering}
	\includegraphics[width=0.6\textwidth]{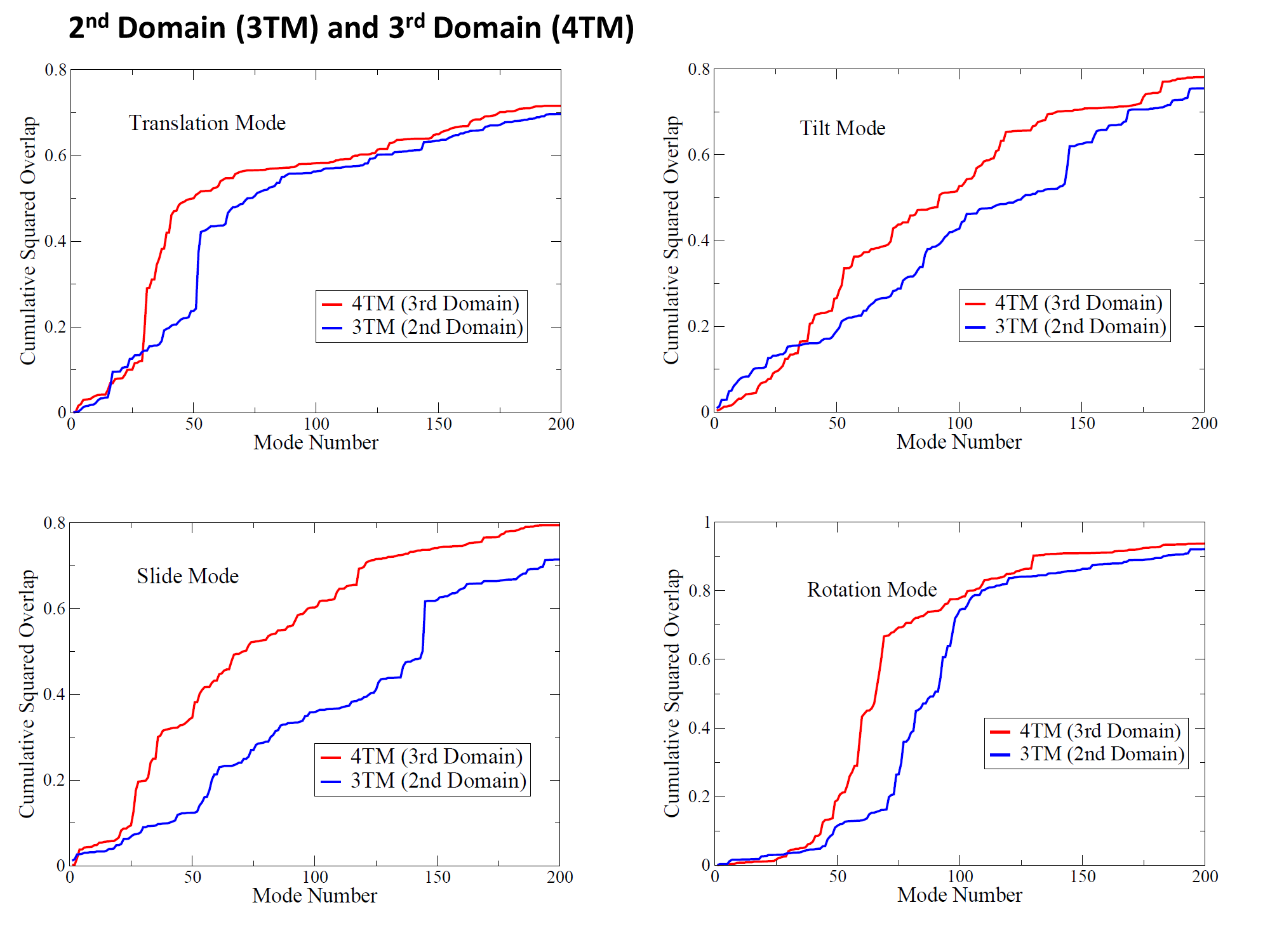}
	\caption{The cumulative squared overlap values for the 2nd TM domain of the 3TM human VKOR model and the 3rd TM domain of the 4TM human VKOR model in translational movement along the axis, rotational movement around axis, slide movement perpendicular to the axis towards/away from the center of the bundle, and tilt perpendicular to the axis away from the center of bundle.}
	\label{fig:6 normal mode analysis 4}
	\end{centering}
\end{figure}
\begin{figure}
	\begin{centering}
	\includegraphics[width=0.6\textwidth]{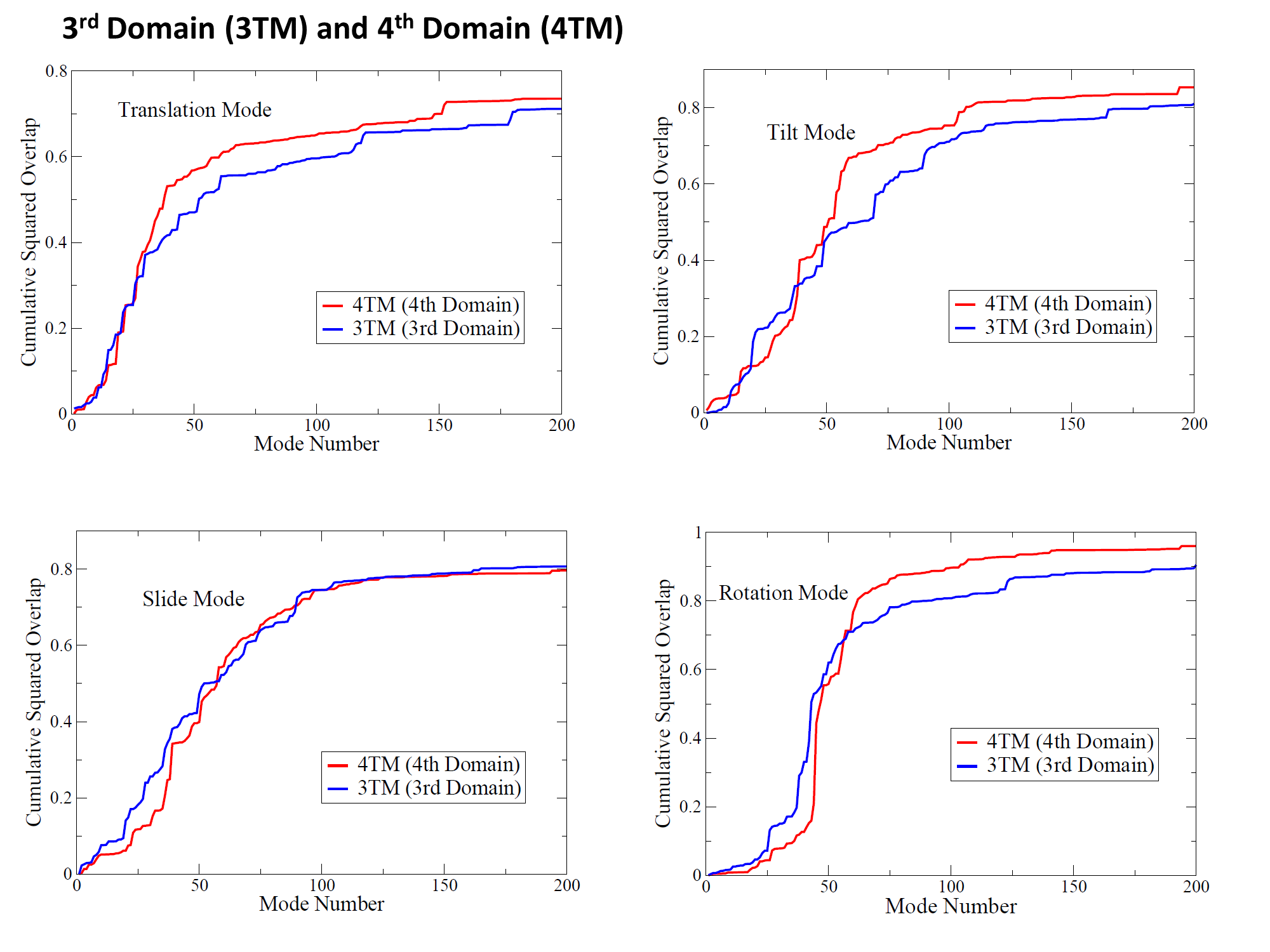}
	\caption{The cumulative squared overlap values for the 3rd TM domain of the 3TM human VKOR model and the 4th TM domain of the 4TM human VKOR model in translational movement along the axis, rotational movement around axis, slide movement perpendicular to the axis towards/away from the center of the bundle, and tilt perpendicular to the axis away from the center of bundle.}
	\label{fig:7 normal mode analysis 5}
	\end{centering}
\end{figure}
\begin{figure}
	\begin{centering}
	\includegraphics[width=0.4\textwidth]{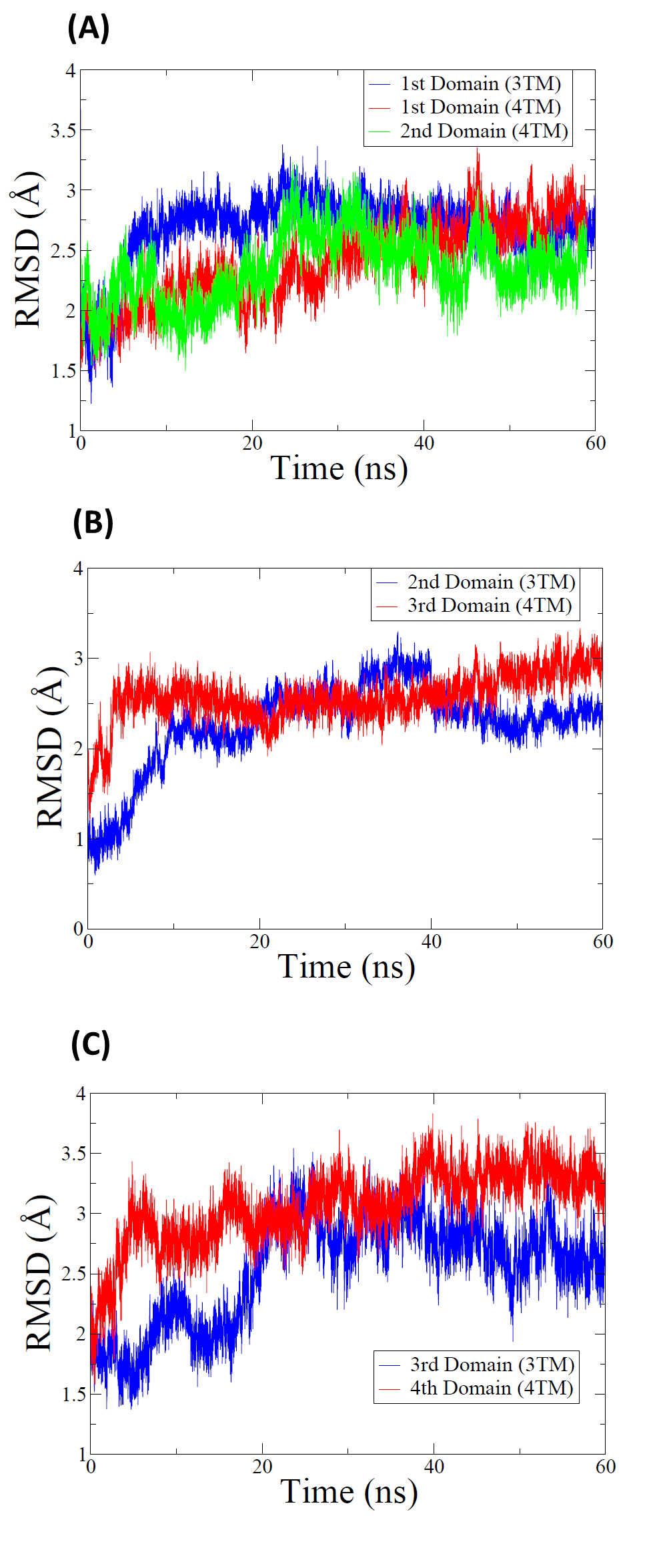}
	\caption{The backbone RMSD (Root Mean Squared Deviation) values of each domain for the 3TM/4TM human VKOR during 60 ns molecular dynamics simulation. The loop regions are excluded in the calculation of backbone RMSD value. The RMSD values for the (A) 1st TM domains of the 3TM/4TM human VKOR models. The backbone RMSD value of the 2nd TM domain of the 4TM human VKOR model is also shown (green). (B) 2nd TM domain of the 3TM human VKOR and its corresponding 3rd TM domain of the 4TM human VKOR model (C) 3rd TM domain of the 3TM human VKOR and its corresponding 4th TM domain of the 4TM human VKOR model. }
	\label{fig:8 molecular dynamics}
	\end{centering}
\end{figure}
\begin{figure}
	\begin{centering}
	\includegraphics[width=0.6\textwidth]{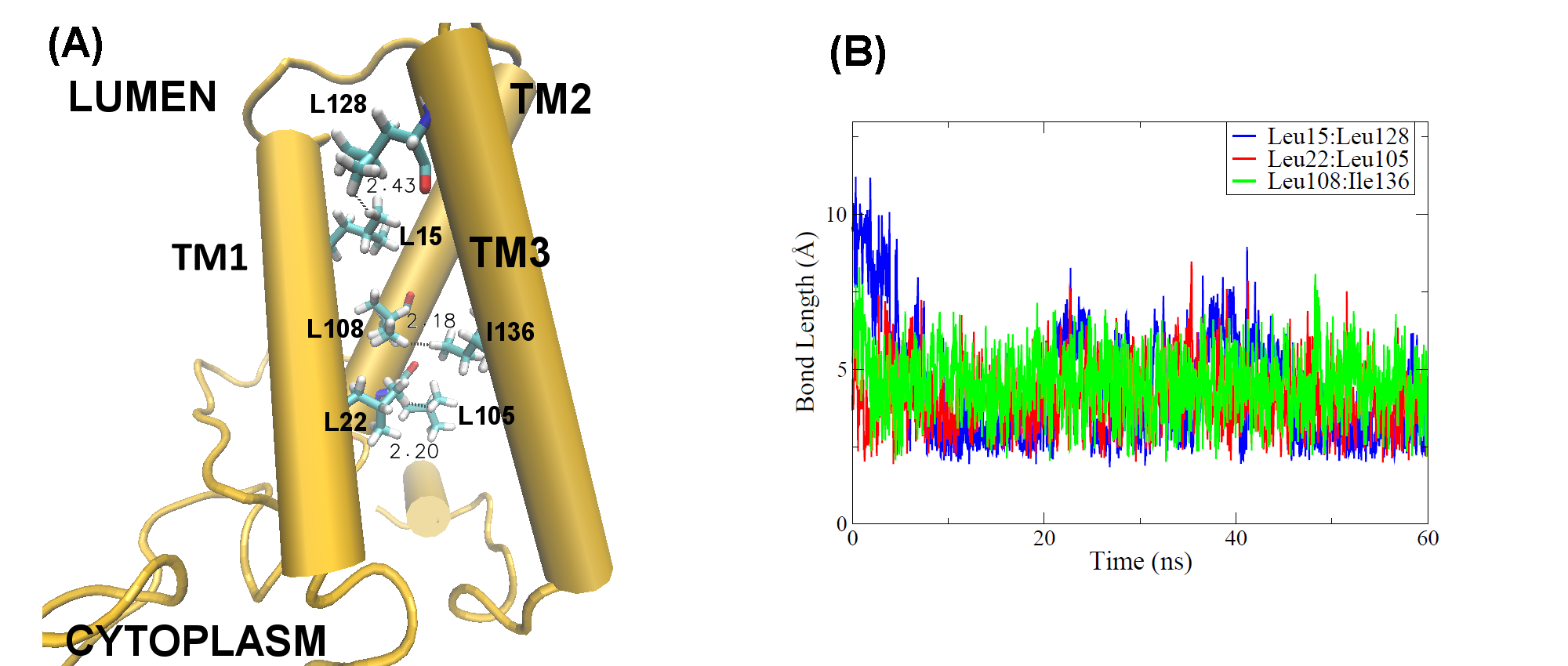}
	\caption{(A) Leucine/Isoleucine residues participating in the hydrophobic interactions. It contributes to the stability of 3TM domains in the 3TM human VKOR model. The numbers are distances in \AA. (B) The bond distances between the leucine/isoleucine residues measured during 60 ns simulation. The bond lengths are measured for the specific hydrogens belong to the leucine/isoleucine residues. The bond lengths do not correspond to the closest distances between the residues.}
	\label{fig:9 leucine/isoleucine}
	\end{centering}
\end{figure}
For the last snapshots of the dynamically generated human 3TM/4TM VKOR models through 60 ns MD simulation (Figure 2) \cite{Wu2}, we investigated the mobility of the TM domains employing normal mode analysis. This approach is effective for understanding the mobility of individual TM helices for a) translational movement along the axis b) rotational movement around axis c) slide movement perpendicular to the axis towards/away from the center of the bundle d) tilt perpendicular to the axis away from the center of bundle. The four different mobility modes are shown for the 4TM human VKOR model (Figure 3) and the 3 TM human VKOR model (Figure 4).
For the human 4TM VKOR model (Figure 3), the 4th TM domain including CxxC motif (including C132 and C135) shows large mobility in the translation, rotation, slide and tilt movements in relatively small mode number (50-100 approximately). Interestingly, the mobility of the 1st TM domain increases in translation, slide and tilt movements in high frequency mode (mode number 150-200 approximately).
For the human 3TM VKOR model (Figure 4), the 3rd TM domain including CxxC motif shows large mobility in translation, rotation, slide and tilt movements in low frequency mode. It is a common feature that the TM domain including CxxC motif shows significant mobility for both human 3TM/4TM VKOR models.
We compare the cumulative squared overlap values for the 1st TM domains of the 3TM/4TM human VKOR models (Figure 4), 2nd TM domain of the 3TM human VKOR and 3rd TM domain of the 4TM human VKOR (Figure 5), and 3rd TM domain of the 3TM human VKOR and 4th TM domain of the 4TM human VKOR  (Figure 6). The mobility of the 1st TM domain of the 4TM human VKOR model shows more mobility than that of the 1st domain of the 3TM human VKOR model in the tilt, slide, and rotation mode (Figure 4). In a similar way, the mobility of the 4th TM domain of the 4TM human VKOR model shows more mobility than that of its corresponding 3rd domain of the 3TM human VKOR model in the tilt, rotation and translation mode (Figure 6). Significantly, the mobility of the 3rd TM domain of the 4TM human VKOR model is greater than that of its corresponding 2nd domain of the 3TM human VKOR model in the translation, tilt, slide, and rotation mode (Figure 5). NMA analysis shows that the mobility of the individual TM domain of the 4TM human VKOR is more than that of the its corresponding TM domain of the 3TM human VKOR model.
We compare the backbone RMSD values of each domain (excluding loop regions) for the 3TM/4TM human VKOR models during 60 ns simulation (Figure 7). The backbone RMSD values show large fluctuation up to 20 ns for all the TM domains for both 3TM/4TM human VKOR models, implying that the dynamical equilibration process occurs up to 20 ns in the simulation. After 20 ns, backbone RMSD of all the TM domains reach plateau values, implying that the system approaches to equilibrium state.
The less mobility of the 3TM human VKOR model is relevant to hydrophobic interactions between the leucine/isoleucine residues on each TM domain. The hydrophobic interactions between the leucine/isoleucine residues contributes to the stability of each TM domain: Leu22 (TM1):Leu105(TM2), Leu15(TM1):Leu128(TM3), and Leu108(TM2):Ile136(TM3) \cite{sangwook}. Figure 9 shows the hydrophobic interactions between leucine/isoleucine residues and their bond lengthes measured during 60 ns simulation. Up to 20 ns simulation, the bond lengths show dramatic changes, implying the dynamical equilibration process.
\section{CONCLUSIONS}
The NMA analysis based on the elastic network model estimates the mobility of the TM helices in the translation, rotation, slide and tilt mode for the last snapshot of the 3TM /4TM VKOR models. In an alternative approach to NMA, the backbone RMSD value (excluding loop regions) from the MD simulation trajectory shows the entire time profile of individual TM domain movement during 60 ns. Interestingly, the NMA analysis results are compatible with the backbone RMSD results after 40 ns of the MD simulation. From the analysis of the TM domain mobility based on the NMA and MD simulation, we conclude that the TM domains of the 4TM human VKOR model are more mobile than their corresponding TM domains of the 3TM human VKOR model. It implies the 3TM human VKOR model is more stable than the 4TM human VKOR model. The origin of the stability of the 3TM human VKOR model comes from the hydrophobic interactions between the leucine/isoleucine residues on each TM domain. The movement of individual TM domain of the 3TM human VKOR model is more suppressed by the hydrophobic interactions between the leucine/isoleucine residues than that of the 4TM human VKOR model.
\begin{acknowledgments}
The authors would like to thank Dr. Lee G. Pedersen for helpful discussions.
This work was supported by a Research Grant of Pukyong National University (2014).
\end{acknowledgments}

\end{document}